\documentclass[aps,preprint,amsmath,amssymb,superscriptaddress,nofootinbib]{revtex4}
\usepackage{multirow}
\usepackage{graphicx}
\usepackage{mathtools}
\usepackage[makeroom] {cancel}
\usepackage{caption}

\begin{document}

\title{Searching for the Leptophilic Gauge Boson Z$_{l}$ at Future $e^{+}e^{-}$ Colliders} 

\author{S. O. Kara}
\email[]{seyitokankara@gmail.com} 
\affiliation{Niğde Ömer Halisdemir University, Bor Vocational School, 51240, Niğde, Türkiye}

 \vspace{0.5cm}

\begin{abstract}

We study the discovery prospects of the leptophilic photon $Z_{l}$, a hypothetical gauge boson beyond the Standard Model, at future electron–positron colliders. Our analysis is based on the process $e^+e^- \to \mu^+\mu^-$ and consistently includes realistic effects such as initial state radiation and beamstrahlung. Using the benchmark parameters of FCC-ee, CEPC, ILC, and CLIC, we demonstrate the complementarity of circular and linear colliders: circular machines achieve excellent sensitivity to very small couplings, down to $g_\ell \sim 10^{-4}$, while high-energy linear colliders extend the search reach into the TeV domain. 

\end{abstract}

\maketitle

\textbf{1. Introduction}

Proton decay into an electron and a photon ($p \to e\gamma$) is strictly forbidden by baryon and lepton number conservation. The idea of gauging these conserved numbers, namely baryon and lepton numbers, in analogy with the electric charge in QED, was first suggested by Lee and Yang in 1955 \cite{1} and by Okun in 1969 \cite{2}, respectively. This fascinating particle has been the subject of many studies in the literature \cite{3,4,5,6,7,8,9,10,11,12,13,14,15,16,17}. Later works introduced a massive variant, denoted $Z_\ell$, as a hypothetical new gauge boson \cite{18,19,20,21}. The defining feature of this boson is that its interactions with quarks are suppressed or absent, while it couples more strongly to leptons.

In this paper, we study the dynamics of this hypothetical leptophilic photon at the next generation of $e^+e^-$ colliders: the International Linear Collider (ILC), the Compact Linear Collider (CLIC), the Future $e^{+}$$e^{-}$ Circular Collider
(FCC-ee) and the Circular Electron Positron Collider (CEPC). The initial center-of-mass energy of the ILC is proposed to be 250~GeV, with a planned upgrade to 500~GeV \cite{22,23}. The CLIC is designed to operate in three stages: 380, 1500, and 3000~GeV \cite{24,25}. The FCC-ee aims to operate at 240~GeV with extremely high luminosity \cite{26}, while the CEPC is also planned to run at 240~GeV \cite{27,28}. Table 1 summarizes the key beam parameters of these colliders. Here, E represents the center-of-mass energy, L signifies the integrated luminosity, N denotes the number of particles in the bunch, ${\sigma}_{x}$ and ${\sigma}_{y}$ are the RMS transverse beam sizes at Interaction Points (IP) and ${\sigma}_{z}$ is the RMS bunch length. 

The search for new neutral gauge bosons remains a central task in current and future colliders, as such particles appear in many extensions of the Standard Model (SM). Leptophilic gauge bosons are particularly well-motivated because they can evade stringent hadron collider bounds by coupling predominantly to leptons. In this study, we focus on the $\mu^+\mu^-$ final state: the $e^+e^-$ channel suffers from an overwhelming Bhabha background, while the $\tau^+\tau^-$ channel is complicated by missing energy from $\tau$ decays. Thus, the dimuon channel provides the cleanest and most sensitive probe of a leptophilic boson at $e^+e^-$ colliders. Throughout this work, we consider the signal process $e^{+}$$e^{-}$$\rightarrow$$\gamma$,$Z$,$Z_{l}$$\rightarrow$$\mu^{+}$$\mu^{-}$ and the background process $e^{+}$$e^{-}$$\rightarrow$$\gamma$,$Z$$\rightarrow$$\mu^{+}$$\mu^{-}$. These are simulated using the \textsc{CalcHEP} framework\cite{29,30}, including realistic effects such as Initial State Radiation (ISR) and Beamstrahlung (BS)\cite{31,32}. The collider parameters used for ISR and BS are listed in Table 1.

Compared with earlier works, our study provides the first systematic comparison of the discovery potential of $Z_\ell$ across all major proposed $e^+e^-$ colliders, consistently incorporating ISR and BS effects. This comprehensive approach enables a more accurate evaluation of collider sensitivity and highlights their complementarity in probing both low- and high-mass regions of the parameter space.

\begin{table} [h!]
\centering
\textit{Table 1: Key beam parameters of future $e^+e^-$ colliders.}

\vspace{0.5cm}

\label{tab1}
\begin{tabular}{|c|c|c|c|c|c|c|c|}
\hline

\hline

\hline
\textbf {\textit{Parameters}}    &  \textbf {\textit{FCC-ee}}  & \textbf {\textit{CEPC}}  & \textbf {\textit{ILC-1}}  & \textbf {\textit{ILC-2}} & \textbf {\textit{CLIC-1}} & \textbf {\textit{CLIC-2}} & \textbf {\textit{CLIC-3}}\\
\hline

E($\sqrt{s}$) GeV & 240 & 240 & 250  & 500 & 380 & 1500 & 3000\\
\hline

L($10^{5}{pb}^{-1}$) & 17 & 6 & 1.35  & 1.8 & 1.5 & 3.7 & 5.9\\
\hline

N($10^{9}$) & 18 & 15 & 2 & 2 & 0.52 & 0.37 & 0.37 \\
\hline

${\sigma}_{x}$(nm) & 13.7 & 20.9 & 0.516 & 0.474 & 0.149 & 0.06 & 0.04\\
\hline
${\sigma}_{y}$(nm) & 36 & 60 & 7.66 & 5.86 & 2.9 & 1.5 & 1\\
\hline
${\sigma}_{z}$(${\mu}$m) & 5.3 & 4.4 & 0.3 & 0.3 & 0.07 & 0.044 & 0.044\\
\hline

\end{tabular}
\end{table}

 \vspace{1.0cm}

\textbf{2. The Model}\\

To gauge the leptonic quantum number, we extend the Standard Model (SM) gauge group
\[
SU_C(3) \times SU_W(2) \times U_Y(1)
\]
by an additional abelian symmetry $U'_\ell(1)$. With the experimental discovery of neutrino oscillations\cite{33}, it is no longer valid to conserve electron, muon, and tau lepton numbers individually. Instead, we introduce a single universal lepton charge, common to $e$, $\mu$, $\tau$, and their corresponding neutrinos. 

In this framework, the free-field Lagrangian is modified as
\begin{equation}
\partial_\mu \;\to\; D_\mu = \partial_\mu - i g_2 \textbf{T} \cdot \textbf{A}_\mu - i g_1 \frac{Y}{2} B_\mu - i g_\ell a_\ell B'_\mu ,
\label{eq:covariant}
\end{equation}
where $g_2$, $g_1$, and $g_\ell$ are the interaction constants, $\textbf{T}$ is the isospin operator of the relevant fermion or Higgs multiplet, $Y$ is hypercharge, and $a_\ell$ is the lepton charge of the multiplet. The gauge fields are $\textbf{A}_\mu$, $B_\mu$, and $B'_\mu$. A mechanism is required to provide mass to the leptophilic boson $Z_\ell$, associated with the $B'_\mu$ field. For this purpose, we introduce a scalar Higgs field $\Phi$ carrying lepton charge. The resulting interaction Lagrangian, invariant under 
\[
SU_C(3) \times SU_W(2) \times U_Y(1) \times U'_\ell(1),
\]
takes the form
\begin{equation}
\mathcal{L} = \mathcal{L}_{SM} + \mathcal{L}' ,
\end{equation}
where $\mathcal{L}_{SM}$ is the Standard Model Lagrangian, and
\begin{equation}
\mathcal{L}' = -\frac{1}{4}F'_{\mu\nu}F^{\mu\nu'} + g_\ell J^\mu_{\text{lep}} B'_\mu 
+ (D_\mu \Phi)^\dagger (D^\mu \Phi) + \mu^2 |\Phi|^2 - \lambda |\Phi|^4 .
\label{eq:Lagrangian}
\end{equation}
Here,
\begin{equation}
F'_{\mu\nu} = \partial_\mu B'_\nu - \partial_\nu B'_\mu
\end{equation}
is the field strength tensor, and
\begin{equation}
J^\mu_{\text{lep}} = \sum_\ell a_\ell \left( \bar{\nu}_\ell \gamma^\mu \nu_\ell + \bar{\ell} \gamma^\mu \ell \right)
\end{equation}
is the leptonic current interacting with $Z_\ell$. The scalar $\Phi$ is a singlet complex Higgs field that provides mass to $Z_\ell$ after spontaneous symmetry breaking.

The parameter space of the model is constrained by precision electroweak data. Previous studies~\cite{34,35} have derived limits on generic $Z'$ bosons. For $Z_\ell$, we adopt the bound~\cite{34}
\begin{equation}
\frac{M_{Z_\ell}}{g_\ell} \;\gtrsim\; 7~\text{TeV}.
\label{eq:bound}
\end{equation}
This relation restricts the allowed values of the coupling $g_\ell$ for a given boson mass $M_{Z_\ell}$. Table 2 lists the corresponding upper bounds on $g_\ell$ for different $M_{Z_\ell}$ values. 

This mass-to-coupling ratio plays a central role in characterizing the properties of the leptophilic boson and constraining the parameter space for collider searches.

\vspace{0.5cm}

\begin{table} [h!]
\textit{Table 2: Upper bounds of $g_{l}$ for different values of Z$_{l}$ mass.}


\label{tab2}

\begin{tabular}{|c |c|c | c|c | c|c | c|c | c|c|}
\hline
 $M_{Z_{l}}$(TeV) &  0.24 & 0.25 & 0.38 & 0.5 & 0.24 & 1.0 & 1.5 & 2.0 & 2.5 & 3.0 \\
 \hline
 $g_{l}$ & 0.033 & 0.25 & 0.035 & 0.05 & 0.07 & 0.14 & 0.21 & 0.28 & 0.35 & 0.42  \\
 \hline
\end{tabular}
\end{table}


\textbf{3. Production of the Z$_{l}$ boson at the electron positron colliders}


Future $e^+e^-$ colliders are designed to provide extremely high luminosities and precision measurements, not only for Standard Model (SM) particles such as the Higgs boson but also for exploring physics beyond the SM. The leptophilic boson $Z_\ell$, introduced through the additional $U'_\ell(1)$ gauge symmetry, represents an intriguing candidate for new physics at these facilities. 

The Future Circular Collider (FCC), a particle accelerator complex proposed by CERN, is being designed to push the frontiers of particle physics beyond the capabilities of the Large Hadron Collider (LHC). The FCC would operate in two stages: FCC-ee, a precision factory for producing Higgs bosons, Z bosons, W bosons, and top quarks; and FCC-hh, which will directly search for new, heavy particles and phenomena at the highest energy scale ever achieved. The FCC-ee, as an electron-positron collider \cite{36,37}, is the perfect machine to complement searches for new particles that only interact with leptons. Its clean, high-luminosity environment and precise control over collision energy make it well-suited for discovering these kinds of particles. The paper \cite{38} considers the discovery opportunities of the leptophilic Z$^{'}$ bosons at FCC-ee. The analysis in this paper   shows that the FCC-ee has significant discovery potential for Z$^{'}$.

In this work, we analyze the process $e^+e^- \to \mu^+\mu^-$ in the presence of a leptophilic gauge boson $Z_\ell$. Simulations are performed using the \textsc{CalcHEP} framework, with ISR and BS consistently included. The $\mu^+\mu^-$ final state is chosen because:
\begin{itemize}
    \item the $e^+e^-$ channel suffers from overwhelming Bhabha scattering background,
    \item the $\tau^+\tau^-$ channel is complicated by missing neutrinos from $\tau$ decays,
    \item the dimuon final state provides the cleanest and most sensitive probe.
\end{itemize}

Figure 1 shows the total cross section $\sigma(e^+e^- \to \mu^+\mu^-)$ as a function of the center-of-mass energy ($\sqrt{s}$) for FCC-ee, CEPC, ILC, and CLIC. The cross section decreases with energy, being a few pb around $\sqrt{S}\approx 200~\text{GeV}$ and dropping to $\mathcal{O}(10^{-1})$ pb at $\sqrt{S}\approx 500~\text{GeV}$.

Figure 2 presents the cross section as a function of the leptophilic boson mass $M_{Z_\ell}$ for different collider energies. Circular colliders such as FCC-ee and CEPC at 240~GeV, as well as the ILC at 250~GeV, provide excellent sensitivity in the low-mass region near resonance. Higher-energy machines like CLIC (380~GeV) and ILC (500~GeV) extend the accessible mass range, maintaining measurable cross sections up to several hundred GeV.

The resonance behavior is illustrated in Figure 3, where the FCC-ee cross section is plotted at $\sqrt{s} = 240$~GeV for different coupling strengths $g_\ell$. A sharp resonance peak is observed near $M_{Z_\ell} \approx \sqrt{s}$, with the cross section reaching values orders of magnitude larger than the SM prediction. Even for $g_\ell$ as small as $10^{-3}$, the signal remains distinguishable from the SM background.

Figure 4 complements this by fixing $M_{Z_\ell}=240$~GeV and varying the collider energy. The resonance is maximized when $\sqrt{s}=M_{Z_\ell}$, while away from resonance the cross section falls toward the SM expectation.

ISR and BS significantly affect the resonance structure. Figure 5 shows the production cross section at $\sqrt{s}=240$~GeV and $g_\ell=0.03$, comparing the idealized signal with and without ISR/BS against the SM background. The inclusion of ISR and BS reduces and broadens the resonance peak due to the effective collision energy spread.

Figure 6 illustrates the dependence of the cross section on $g_\ell$ at FCC-ee with $M_{Z_\ell} = 240$~GeV. While ISR reduces the normalization, the overall $g_\ell$ dependence remains unchanged. Including BS further suppresses the signal by about one order of magnitude across the range. 

Differential distributions provide an additional probe of $Z_\ell$. Figure 7 shows $d\sigma/dM_{\mu^+\mu^-}$ at FCC-ee for $g_\ell=0.03$ and several $M_{Z_\ell}$ values. Sharp resonance peaks are clearly visible above the SM background, with positions corresponding to $M_{Z_\ell}$.

Figure 8 fixes $M_{Z_\ell}=200$~GeV and varies $g_\ell$, confirming the expected quadratic scaling of the resonance peak with coupling.

To quantify discovery potential, we define the statistical significance
\begin{equation}
S = \frac{\sigma_{\text{signal}} - \sigma_{\text{SM}}}{\sqrt{\sigma_{\text{SM}}}} \, \sqrt{L_{\text{int}}},
\end{equation}
where $\sigma_{\text{signal}}$ and $\sigma_{\text{SM}}$ are the signal and background cross sections, and $L_{\text{int}}$ is the integrated luminosity. Events are selected using the cuts
\[
|M_{\mu^+\mu^-} - M_{Z_\ell}| < 10~\text{GeV}, \qquad |\eta_\mu| < 2.
\]

Figures 9 -- 12 summarize the discovery ($5\sigma$) and observation ($3\sigma$) reaches in the $(M_{Z_\ell}, g_\ell)$ plane for FCC-ee, CEPC, ILC, and CLIC. Circular colliders probe very small couplings ($g_\ell \sim 10^{-4}$) in the low-mass region, while linear colliders extend the reach to the TeV scale.

 \vspace{2.5cm}
 
\textbf{5. Conclusions}\\

In this work, we have investigated the discovery potential of the leptophilic gauge boson $Z_\ell$ at future electron--positron colliders, namely FCC-ee, CEPC, ILC, and CLIC. The analysis was based on the process $e^+e^- \to \mu^+\mu^-$, with Initial State Radiation (ISR) and Beamstrahlung (BS) consistently included through the \textsc{CalcHEP} framework.

Our results demonstrate the complementarity of circular and linear collider designs. Circular machines (FCC-ee and CEPC) are particularly sensitive to very small couplings, reaching values as low as $g_\ell \sim 10^{-4}$ in the low-mass regime. High-energy linear colliders (ILC and CLIC) extend the reach into the multi-hundred GeV and TeV domains, maintaining measurable sensitivity for heavier states. Among these, FCC-ee provides the strongest reach at low masses due to its extremely high luminosity and clean experimental environment.

Compared to previous studies, which typically focused on individual colliders or neglected beam-related effects, this work presents the first systematic comparison of $Z_\ell$ discovery potential across all major proposed $e^+e^-$ colliders, with ISR and BS effects consistently taken into account. This comprehensive approach provides a robust evaluation of collider sensitivities and highlights the complementarity of different collider designs in exploring the parameter space of leptophilic interactions.

Future extensions of this research could incorporate detailed detector simulations, systematic uncertainties, and additional observables or kinematic variables that may further enhance sensitivity. Moreover, investigating possible connections between the leptophilic photon and other beyond-Standard-Model scenarios---for example, its interplay with dark sector mediators---could broaden the scope of this study and further guide new physics searches at upcoming high-luminosity colliders.

\begin{figure}[h!]
    \centering
    \includegraphics[width=0.8\textwidth]{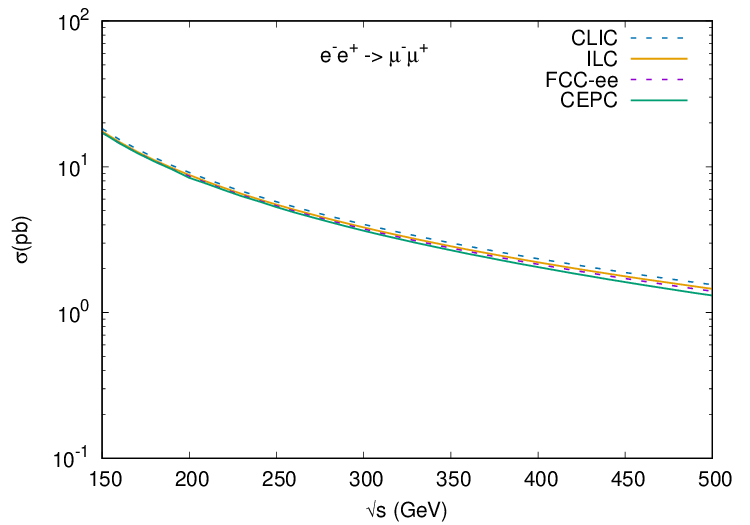}
    
    {\textit{Figure 1: Cross section of $e^+e^- \to \mu^+\mu^-$ as a function of $\sqrt{s}$ for future colliders (FCC-ee, CEPC, ILC, and CLIC).}}
    \label{fig:tekli}
\end{figure}

\begin{figure}[h!]
    \centering
    \includegraphics[width=0.8\textwidth]{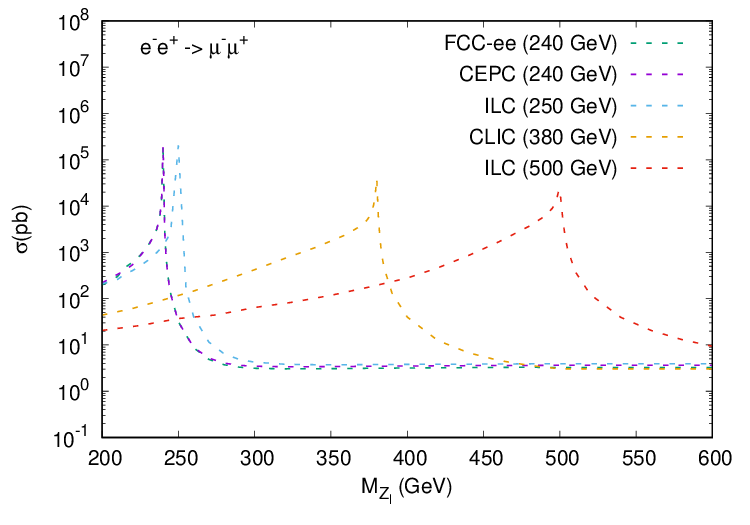}
    
    \textit{Figure 2: Cross section of $e^+e^- \to \mu^+\mu^-$ as a function of $M_{Z_\ell}$ for different collider energies.}
    \label{fig:tekli}
\end{figure}

\begin{figure}[h!]
    \centering
    \includegraphics[width=0.8\textwidth]{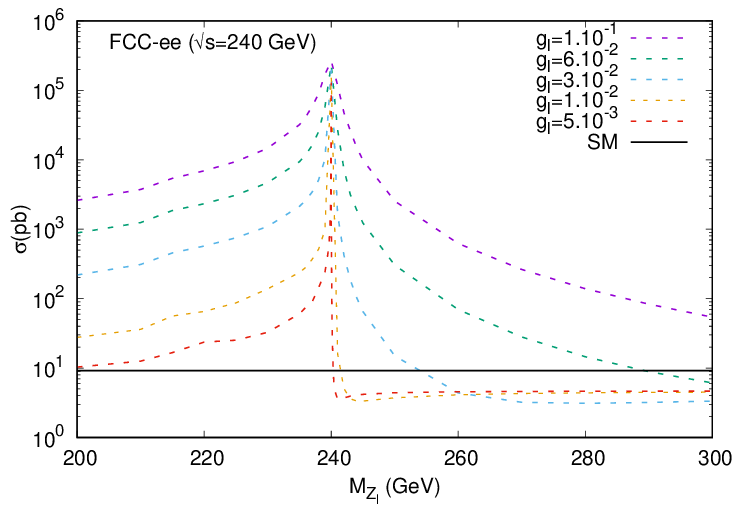}
    
    {\textit{Figure 3: Cross section of $e^+e^- \to \mu^+\mu^-$ at FCC-ee with $\sqrt{s}=240$ GeV as a function of $M_{Z_\ell}$ for different coupling values $g_\ell$, compared with the SM background.}}
    \label{fig:tekli}
\end{figure}

\begin{figure}[h!]
    \centering
    \includegraphics[width=0.8\textwidth]{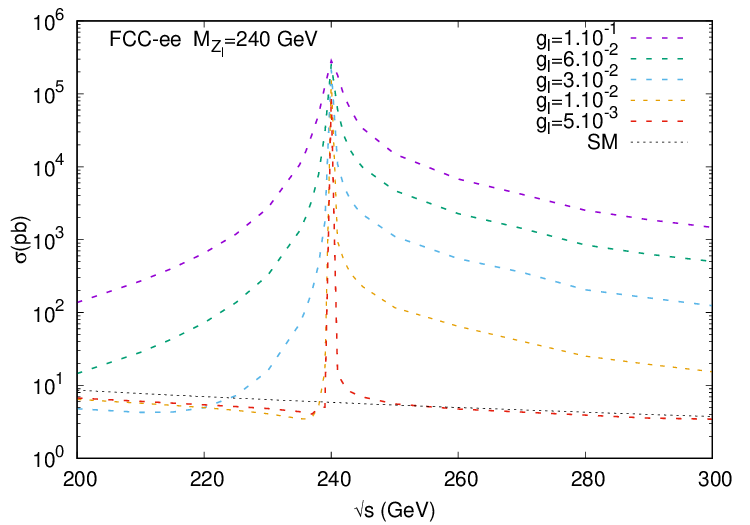}
    
    {\textit{Figure 4: Cross section of $e^+e^- \to \mu^+\mu^-$ at FCC-ee for $M_{Z_\ell}=240$~GeV as a function of $\sqrt{s}$, shown for different $g_\ell$ values.}}
    \label{fig:tekli}
\end{figure}

\begin{figure}[h!]
    \centering
    \includegraphics[width=0.8\textwidth]{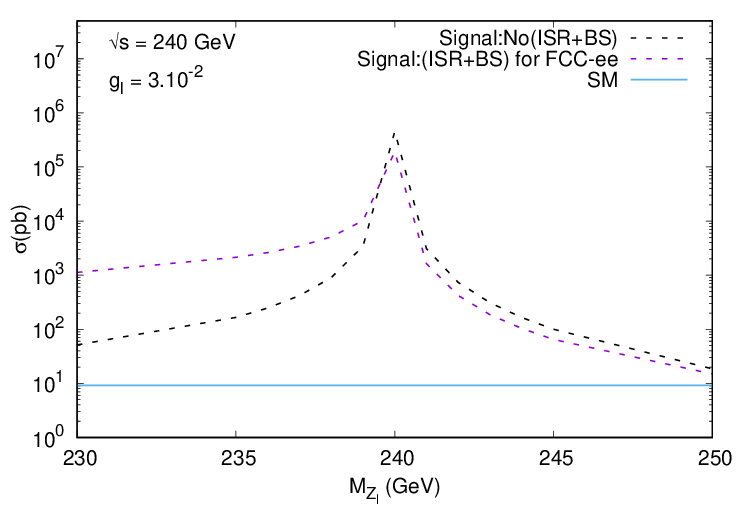}
    
    {\textit{Figure 5: Cross section of $e^+e^- \to \mu^+\mu^-$ at FCC-ee ($\sqrt{s}=240$~GeV, $g_\ell=0.03$), comparing signal with and without ISR+BS against the SM background.}}
    \label{fig:tekli}
\end{figure}

\begin{figure}[h!]
    \centering
    \includegraphics[width=0.8\textwidth]{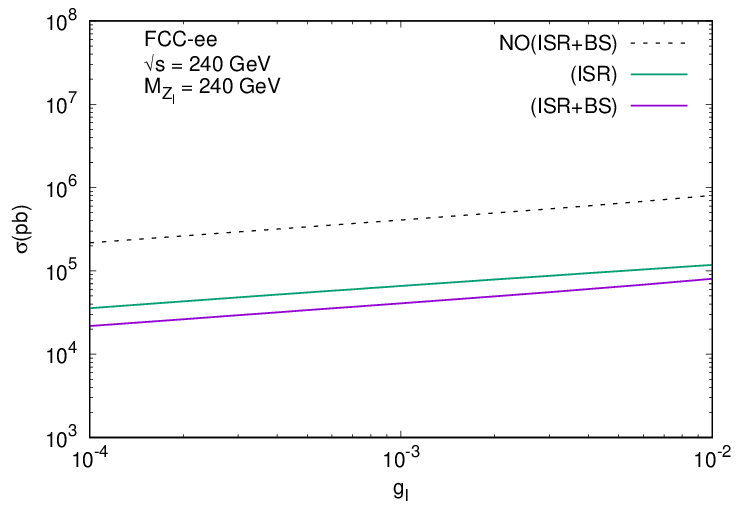}
    
    {\textit{Figure 6: Cross section of $e^+e^- \to \mu^+\mu^-$ at FCC-ee with $\sqrt{s}=240$~GeV and $M_{Z_\ell}=240$~GeV as a function of $g_\ell$, with and without ISR and BS.}}
    \label{fig:tekli}
\end{figure}

\begin{figure}[h!]
    \centering
    \includegraphics[width=0.8\textwidth]{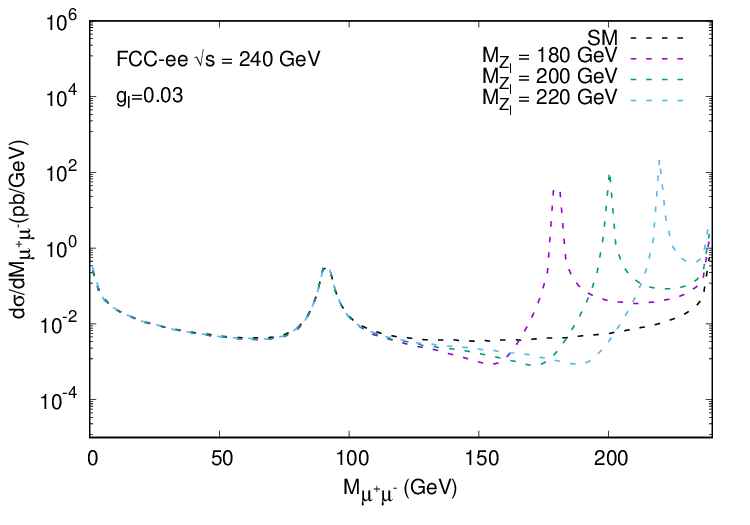}
    
    {\textit{Figure 7: Differential cross section $d\sigma/dM_{\mu^+\mu^-}$ at FCC-ee ($\sqrt{s}=240$~GeV, $g_\ell=0.03$) for different $M_{Z_\ell}$ values, compared with the SM background.}}
    \label{fig:tekli}
\end{figure}

\begin{figure}[h!]
    \centering
    \includegraphics[width=0.8\textwidth]{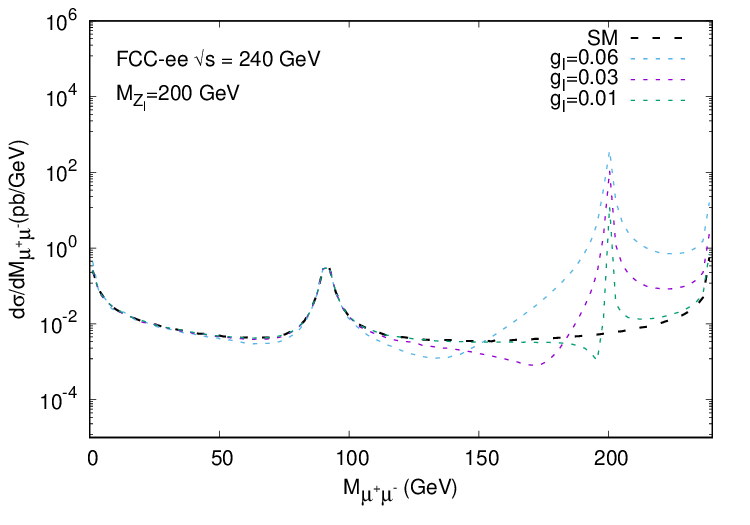}
    
    {\textit{Figure 8: Differential cross section $d\sigma/dM_{\mu^+\mu^-}$ at FCC-ee ($\sqrt{s}=240$~GeV, $M_{Z_\ell}=200$~GeV) for different $g_\ell$ values, compared with the SM background.}}
    \label{fig:tekli}
\end{figure}

\begin{figure}[h!]
    \centering
    \includegraphics[width=0.8\textwidth]{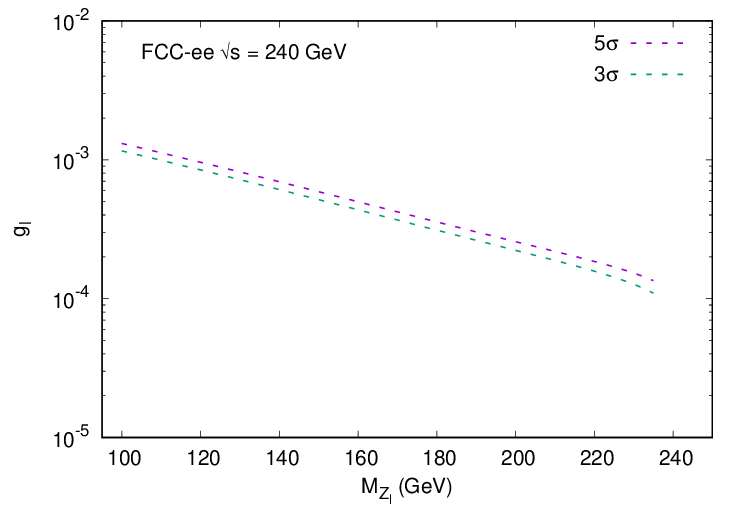}
    
    {\textit{Figure 9: Discovery (5$\sigma$) and observation (3$\sigma$) limits for $Z_\ell$ at FCC-ee ($\sqrt{s}=240$~GeV).}}
    \label{fig:tekli}
\end{figure}

\begin{figure}[h!]
    \centering
    \includegraphics[width=0.83\textwidth]{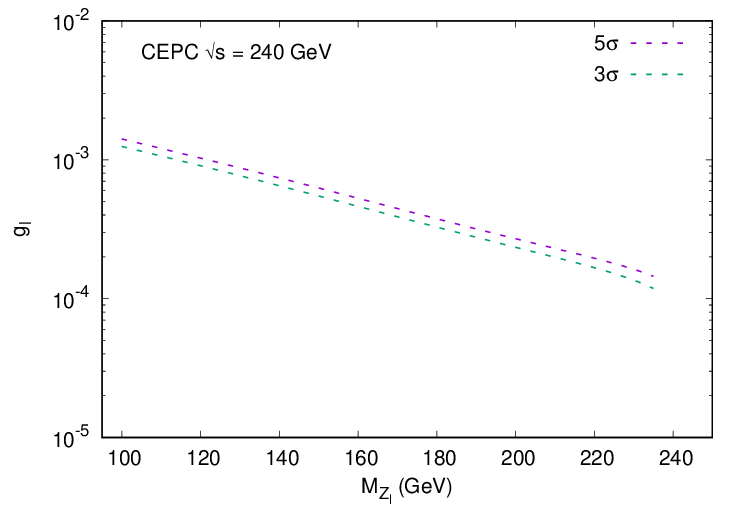}
    
    {\textit{Figure 10: Discovery (5$\sigma$) and observation (3$\sigma$) limits for $Z_\ell$ at CEPC ($\sqrt{s}=240$~GeV).}}
    \label{fig:tekli}
\end{figure}

\begin{figure}[h!]
    \begin{minipage}{0.48\textwidth}
        \centering
        \includegraphics[width=1.1\textwidth]{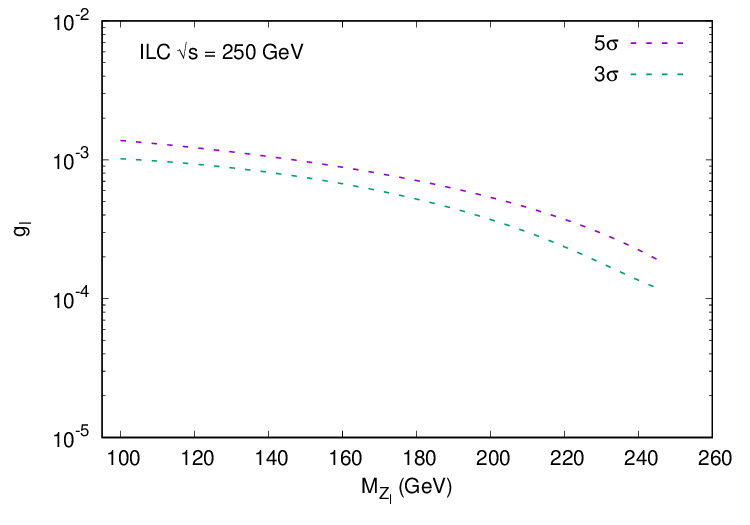}
        
         \text  (a)
        \label{Fig11.eps}
    \end{minipage}\hfill
    \begin{minipage}{0.48\textwidth}
        \centering
        \includegraphics[width=1.1\textwidth]{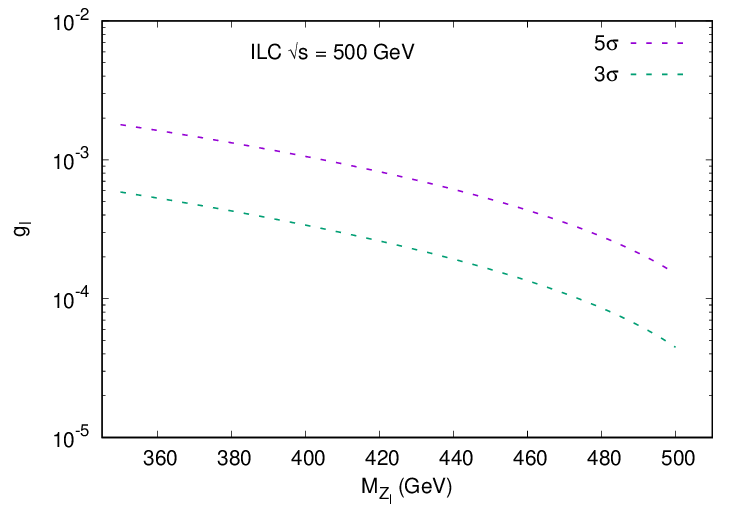}
        
         \text  (b)
        \label{fig:sag}
    \end{minipage}
    
    \vspace{0.5cm}
    
    {\textit{Figure 11: Discovery (5$\sigma$) and observation (3$\sigma$) limits for $Z_\ell$ at the ILC with (a) $\sqrt{s}=250$~GeV and (b) $\sqrt{s}=500$~GeV.}}
    \label{fig:ikili}
\end{figure}

\begin{figure}[h!]
    \centering
    \begin{minipage}{0.48\textwidth}
        \centering
        \includegraphics[width=0.85\textwidth]{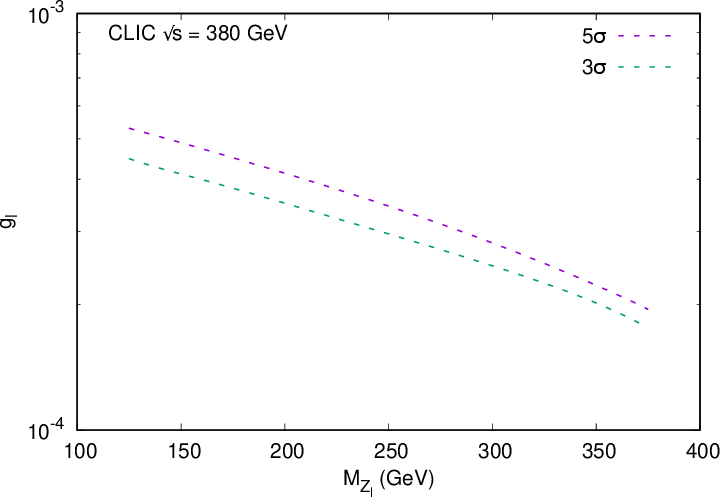}
        
          \text  (a)
        \label{fig:ust_sol}
    \end{minipage}\hfill
    \begin{minipage}{0.48\textwidth}
        \centering
        \includegraphics[width=0.85\textwidth]{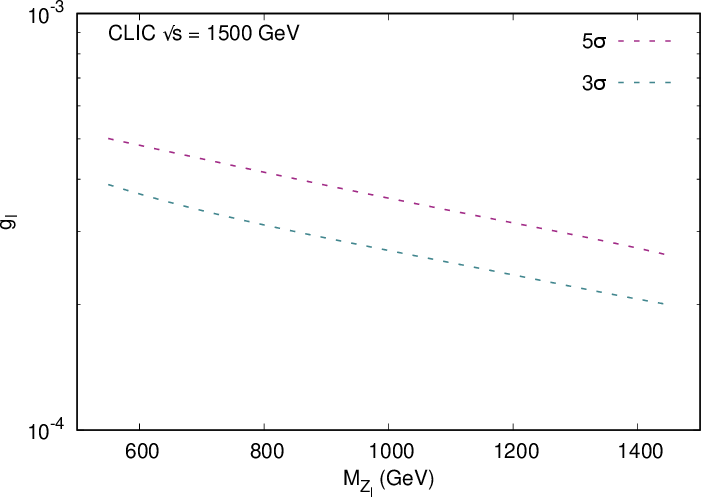} 
        
        \text  (b)
        \label{fig:ust_sag}
    \end{minipage}
    
    \vspace{0.5cm} 

    \begin{minipage}{0.6\textwidth}
        \centering
        \includegraphics[width=0.7\textwidth]{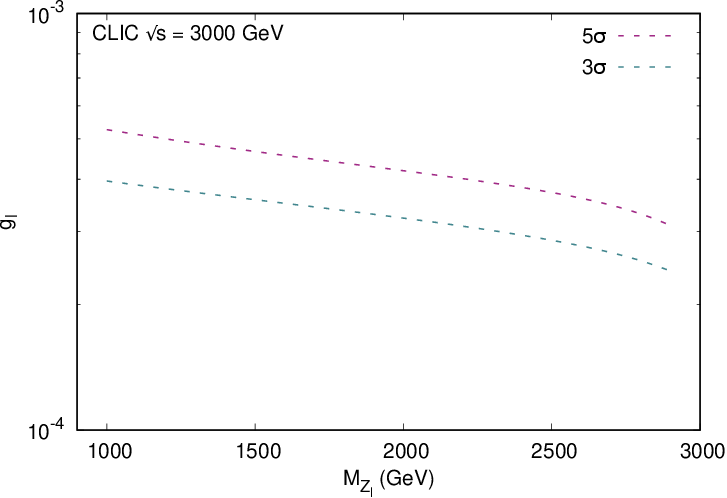}
        
        \text  (c)
        \label{fig:alt}
    \end{minipage} 
    
    \vspace{0.5cm}
    
     \textit{Figure 12: Discovery (5$\sigma$) and observation (3$\sigma$) limits for $Z_\ell$ at CLIC with (a) $\sqrt{s}=380$~GeV, (b) $\sqrt{s}=1500$~GeV, and (c) $\sqrt{s}=3000$~GeV.}
  
    \label{fig:ust_iki_alt_bir}         
\end{figure}

\pagebreak

\newpage

\end{document}